\documentclass[twocolumn,showpacs,amsmath,amssymb,nofootinbib,floatfix,prd]{revtex4}
\usepackage[dvips]{graphicx}
\usepackage{dcolumn}
\usepackage{bm}

\newcommand{\be}{\begin{eqnarray} }
\newcommand{\ee}{\end{eqnarray} }
\newcommand{\beq}{\begin{equation} }
\newcommand{\eeq}{\end{equation} }

\newcommand{\simleq}{\; \raisebox{-0.4ex}{\tiny$\stackrel
{{\textstyle<}}{\sim}$}\;} 

\begin{document} 
\title{
Direct extraction of transversity and its accompanying T-odd distribution 
from the unpolarized and single-polarized Drell-Yan
processes}
       \author{ A.N.~Sissakian}
       \email{sisakian@jinr.ru}
       \author{O.Yu.~Shevchenko}
       \email{shev@mail.cern.ch}
       \author{A.P.~Nagaytsev}
       \email{nagajcev@mail.desy.de}
       \author{O.N.~Ivanov}
       \email{ivon@jinr.ru}
       \affiliation{
       Joint Institute for Nuclear Research,
       141980 Dubna,  Russia
       }

\begin{abstract}
The  Drell-Yan (DY) processes with unpolarized 
colliding hadrons and with the single transversally polarized     
hadron are considered. 
The possibility of direct (without any model assumptions)
extraction of both transversity and its accompanying T-odd parton distribution functions (PDF) is discussed.
For DY processes measurements  planned at GSI 
 the preliminary estimations demonstrate that it is quite real to extract both  transversity and its accompanying T-odd PDF
in the PAX conditions.
\end{abstract}
\pacs{13.65.Ni, 13.60.Hb, 13.88.+e}

\maketitle
The advantage of DY process for extraction of PDF, 
is that there is no need of any fragmentation functions. While the double transversely polarized DY process
$H_1^{\uparrow}H_2^{\uparrow} \to l^{+}l^{-}X$  allows  to directly extract the transversity distributions 
(see ref.\cite{ba} for review),
in the single polarized DY $H_1  H_2^{\uparrow} \to l^{+}l^{-}X$ the access to transversity is rather difficult since 
it enters the respective cross-section  in the complex
convolution with 
another 
unknown T-odd PDF (see below). At the same time 
it is, certainly very desirable to manage to get the transversity PDF from 
unpolarized and single-polarized DY processes
as an alternative possibility.
Besides, T-odd PDF are very intriguing and interesting objects in themselves, 
so that it is very important
to extract them too. 

The main goal of this paper is to investigate the possibility  to 
completely disentangle PDFs corresponding to
the unpolarized and single-polarized DY processes.

Let us first consider the results of ref. \cite{bo1} for both unpolarized and single-polarized DY processes.
 In that paper the Collins-Soper frame\footnote{See \cite{ba} for detail of the respective kinematics.} 
is used ( see Fig. 3 in ref. \cite{bo1}),
where one deals with three angles $\theta$, $\phi$ and $\phi_{S_2}$. Two  
 angles, $\theta$ and $\phi$, are common for both unpolarized and polarized DY processes. 
These are the polar and 
azimuthal angles of lepton pair.
Third angle, $\phi_{S_2}$, does appear when hadron two is transversally polarized, 
and this is just the azimuthal angle of ${\bf S}_{2T}$ measured with respect to lepton plane. 

We consider here the case of pure transverse polarization of hadron two, so that we 
put $\lambda_1=0$  and $|{\bf S}_{1T}|=1 $ ($\lambda_2=0$  and $|{\bf S}_{2T}|=1 $ 
in our notation) in the respective equations of ref. \cite{bo1} 
(Eqs. (21) and  (22) in ref. \cite{bo1})
for unpolarized and single polarized cross-sections.
Besides, taking into account only the dominating electromagnetic contributions 
and neglecting (just as in ref. \cite{bo1})
the ``higher harmonic'' 
term containing  $3\phi$ dependence, one gets the following simplified equations   
for the QPM unpolarized and single-polarized cross-sections :       
\be
\label{unpolcross}
 &  & \frac{d\sigma^{(0)}(H_1H_2\rightarrow l\bar l X)}{d\Omega dx_1 dx_2 d^2 {\bf q}_T}  =  
\frac{\alpha^2}{12 Q^2}\sum\nolimits_{q} e_q^2 \nonumber\\
 &  &\times\Biggl\{(1+{\rm cos}^2\theta)
 {\cal F}[\bar f_{1q} f_{1q}] 
 + {\rm sin}^2\theta {\rm cos}(2\phi)
 \nonumber\\
 &  & \times  {\cal F}
\left[(2\hat{\bf h}\cdot{\bf k}_{1T}\, 
\hat{\bf h}\cdot{\bf k}_{2T}\right.\nonumber\\
 &  &  - \left.  {\bf k}_{1T}\cdot{\bf k}_{2T})\frac{\bar h_{1q}^\perp h_{1q}^\perp}{M_1M_2}
\right]\Biggl\},
\ee
and
\be
\label{e2}
 &  & \frac{d\sigma^{(1)}(H_1 H_2^{\uparrow} \to l\bar l X)}{d\Omega d{\phi}_{S_2} 
dx_1 dx_2 d^2 {\bf q}_T}  = 
\frac{\alpha^2}{12 Q^2}\sum\nolimits_{q} e_q^2\nonumber\\ 
 &  &  \times \Biggl \{ (1+{\rm cos}^2\theta) {\cal F}[\bar f_{1q} f_{1q}]  
  + {\rm sin}^2\theta {\rm cos}(2\phi)\nonumber\\
 &  & \times {\cal F}
\left[(2\hat{\bf h}\cdot{\bf k}_{1T}\, 
\hat{\bf h}\cdot{\bf k}_{2T}
-{\bf k}_{1T}\cdot{\bf k}_{2T})\frac{\bar h_{1q}^\perp h_{1q}^\perp}{M_1M_2}\right] 
 \nonumber \\ 
 &  &  +(1+{\rm cos}^2\theta){\rm sin}(\phi-\phi_{S_2}){\cal F}\left[\hat{\bf h}\cdot{\bf k}_{2T}\frac{\bar f_{1}^q f_{1T}^{\perp q}}{M_2}\right] 
 \nonumber \\
 &  &  -{\rm sin}^2\theta {\rm sin}(\phi+\phi_{S_2}){\cal F}\left[\hat{\bf h}\cdot{\bf k}_{1T}\frac{\bar h_{1q}^\perp h_{1q}}{M_1}\right] 
\Biggl \}.
\ee
Here 
$\hat {\bf h} \equiv {\bf q}_T/|{\bf q}_T |$,
$h_{1q}(x,{\bf k}_T^2)$ is the $k_T$-dependent  transversity distribution,
while  $h_{1q}^\perp (x,{\bf k}_T^2)$  and $f_{1T}^{\perp q}(x,{\bf k}_T^2)$ are $k_T$-dependent T-odd PDFs 
(see ref. \cite{ba} for review).
The convolution product is defined \cite{bo1} as 
\be
\label{e3}
 &  & {\cal F}[\bar f_q f_q]\equiv\int d^2{\bf k}_{1T}
\,d^2{\bf k}_{2T}\,\delta^2({\bf k}_{1T}+{\bf k}_{2T}-{\bf q}_T)\nonumber\\
 &  & \times\left[
f_q(x_1,{\bf k}_{1T}^2)\bar f_q(x_2,{\bf k}_{2T}^2)+(1\leftrightarrow2)\right].
\ee

Let us first consider the purely unpolarized DY process. Notice that Eq. (\ref{unpolcross})
is very inconvenient in application because of the complicated ${q}_{T}$ and ${k}_{T}$ dependence
entering Eq. (\ref{unpolcross}) via the convolution, Eq. (\ref{e3}).
To deal with Eq. (\ref{unpolcross}) the model 
\be
\label{bmodel}
h_{1q}^{\perp}(x,{\bf k}_T^2)=\frac{\alpha_T}{\pi}c_H^q\frac{M_C M_H}{{\bf k}_T^2+M_C^2}{\rm e}^{-\alpha_T{\bf k}_T^2}f_{1q}(x),
\ee
where $M_C=2.3\,GeV$, $c_H^q=1$, $\alpha_T=1\,GeV^{-2}$ and $M_H$ is the hadron mass, was proposed
in ref. \cite{bo1}. With a such assumption one then calculates \cite{bo1,bianconi} 
the coefficient $\kappa\equiv\nu/2$
at $\cos 2\phi$ dependent part of the ratio 
\be
\label{rnohat}
R\equiv \frac{d\sigma^{(0)}/d\Omega}{\sigma^{(0)}},
\ee
which allows to explain\footnote{Notice that the large values of $\nu$ cannot be explained by leading and next-to-leading order perturbative QCD
corrections as well as by the high twists effects (see \cite{bo1} and references therein).}
the anomalous $\cos 2\phi$ dependence \cite{conway,NA10} of the 
unpolarized DY cross-section. However, the author of ref. \cite{bo1} stresses that Eq. (\ref{bmodel})
is just a ``crude model''.
Besides,  Eq. (\ref{bmodel}) can not help us to  extract the quantity
$h_1^\perp$ from the unpolarized DY process. 

Thus, to avoid these problems, let us apply the ${\bf q}_T$ weighting approach
which was first proposed and applied  in refs. \cite{mul1} and \cite{mul2} with respect to 
a particular electron-positron annihilation process and in ref. \cite{mul3} with respect
to semi-inclusive DIS. To use the advantage of ${\bf q}_T$ integration,
one should extract from unpolarized DY process
the properly integrated
over ${\bf q}_T$ ratio (c.f. Eq. (\ref{rnohat}))
\be
\label{rhat}
\hat R=\frac{\int d^2 {\bf q}_T [{\bf |}{\bf q}_T{\bf |}^2/{M_1M_2}][d\sigma^{(0)}/d\Omega]}{\int d^2{\bf q}_T\sigma^{(0)}},
\ee
parametrized as 
\be
\label{rhat1}
\hat R=\frac{3}{16\pi}(\gamma(1+\cos^2\theta)+\hat k\cos 2\phi\sin^2\theta),
\ee
that should be compared\footnote{Obtaining Eg. (\ref{cross}) one sets \cite{bo1}
$\lambda=1$ and $\mu =0$ in the most general equation for $R$ (Eq. (5) in ref. \cite{bo1}),
which is justified \cite{bo1} by the expectation from next-to-leading order QCD  
and the data (refs \cite{conway,NA10}) in the Collins-Soper frame.} 
with the equation  (see refs \cite{bo1,conway}) 
\be
\label{cross}
 &&  R=\frac{3}{16\pi}(1+\lambda\cos^2\theta+ \mu\sin2\theta\cos\phi\nonumber\\
 &&  + (\nu/2)\cos 2\phi\sin^2\theta) 
\quad(\nu \equiv 2\kappa,\lambda\simeq1,\mu\simeq0 ).
\ee

 By virtue of  Eq. (\ref{unpolcross}),
the coefficient $\hat k$ at $\cos 2\phi$ dependent part of $\hat R$ reads 
\be
 &  & \hat k=
\int d^2{\bf q}_T[{\bf| q_T |}^2/M_1M_2]\nonumber\\
 &  & \times\sum_qe_q^2{\cal F} [(2\hat {\bf h}\cdot {\bf k}_{1T}\hat{\bf h}
\cdot {\bf k}_{2T}-{\bf k}_{1T}\cdot{\bf k}_{2T})\frac{\bar h_1^\perp h^\perp_1}{M_1M_2}]\nonumber\\
 &  & \times\left({\int d^2{\bf q}_T \sum_q e_q^2{\cal F}[\bar f_1 f_1]}\right)^{-1},
\ee
and, due to the properly chosen weight $|{\bf q}_T|^2$, the integration
over ${\bf q}_T$ leads\footnote{The normalization condition $\int d^2 {\bf k}_T f_{1q}(x,{\bf k}_T^2)
= f_{1q}(x)$
is used (see, for example ref. \cite{ba}).} to the following simple equation for $\hat k$: 
\be
\label{khat}
\hat k =8\frac{\sum_qe_q^2(\bar h_{1q}^{\perp(1)}(x_1) h_{1q}^{\perp(1)}(x_2)+(1\leftrightarrow2))}
{\sum_qe_q^2(\bar f_{1q}(x_1)f_{1q}(x_2)+(1\leftrightarrow2))},
\ee
where the standard notation \cite{mul1,mul2,mul3}
\be
h_{1q}^{\perp(n)}(x)\equiv\int d^2{\bf k}_T\left(\frac{{\bf k}_T^2}{2M^2}\right)^nh_{1q}^\perp(x,{\bf k}_T^2)
\ee
for the $n$-th moment of ${\bf k}_T$-dependent PDF is used. Thus, one can see that 
the numerator of $\hat k$ is factorized out in the simple product of the first 
moments of $h_1^\perp$ distributions. This allows to directly extract these quantities 
from $\hat k$ which should be measured in unpolarized DY. This, in turn (see below),
allows to directly extract the transversity distributions $h_1$ from the single spin polarized
DY.  Notice that now there is no need in any
model assumptions about ${k}_T$ dependence of  $h_1^\perp$ distributions.

Let us now consider the single transversely polarized DY process  $H_1  H_2^{\uparrow} \to l^{+}l^{-}X$
and define the following single-spin asymmetries (SSA)
\be
\label{ahf}
&&  A_{h(f)}  =  
\nonumber\\
&&\int d\Omega d\phi_{S_2}\sin(\phi\pm\phi_{S_2})[d\sigma({\bf S}_{2T})-d\sigma(-{\bf S}_{2T})]
\nonumber\\
&& \times\left({\int d\Omega d \phi_{S_2}[d\sigma({\bf S}_{2T})+d\sigma(-{\bf S}_{2T})]}\right)^{-1},
\ee
where the single--polarized cross-section is given by Eq. (\ref{e2}).
It is clear that in the difference $d\sigma({\bf S}_{2T})-d\sigma(-{\bf S}_{2T}) $ 
only the terms of Eq. (\ref{ahf}) containing $ {\rm sin}(\phi-\phi_{S_2})$ and 
${\rm sin}(\phi+\phi_{S_2})$
survive (and are multiplied by two). Besides, the properly chosen\footnote{
The analogous weighting procedure was applied \cite{hermes} in the case of transversely polarized SIDIS by the HERMES
collaboration.
} 
weights:  $\sin(\phi+\phi_{S_2})$ and $\sin(\phi-\phi_{S_2})$,  allow to separate the contributions
containing $h_1^\perp$ and $f_{1T}^\perp$ PDF with the result
\be
\label{a1}
A_{h} = -\frac{1}{4}\frac{\sum_{q} e_q^2\,  {\cal F}\left[\frac{\hat{\bf h}\cdot{\bf k}_{1T}}{M_1}\bar h_{1q}^\perp h_{1q}\right]} 
{\sum_{q} e_q^2\,  {\cal F} \left [ \bar f_{1q}f_{1q}\right]},
\ee
and
\be
\label{a2}
A_{f} = \frac{1}{2}  \frac{ \sum_q e_q^2 \,{\cal F}\left [\frac {\hat{\bf h}\cdot {\bf k}_{2T} }{M_2}{\bar f}_{1}^q f_{1T}^{\perp q} \right ]}{ 
\sum_q e_q^2 \,{\cal F} 
\left [\bar f_{1q} f_{1q} \right]}.
\ee
The asymmetries like $A_f$ given by Eqs. (\ref{ahf}), (\ref{a2}) and 
their application with respect to Sivers function 
$f_{1T}^\perp(x,{\bf k}_T^2)\equiv-(M/2|{\bf k}_T|)\Delta^N_{q/H^\uparrow}(x,{\bf k}_T^2)$
extraction from the data were considered in detail in refs. \cite{ans2, efremov}, so that
we concentrate here on the asymmetry $A_h$ given by Eqs. (\ref{ahf}) and (\ref{a1}).

Notice that asymmetry $A_h$ given by Eqs. (\ref{ahf}), (\ref{a1}) is inconvenient
in application because of the complicated ${q}_T$ and ${k}_T$ dependence 
entering the convolution. So, we again apply the ${q}_T$ integration method \cite{mul1,mul2,mul3}
(see also its application for the SIDIS processes in ref. \cite{hermes} 
and for the Sivers PDF  extraction from the single polarized DY in ref. \cite{efremov} ):
\begin{widetext}
\be
\label{aexpfinal}
\hat A_{h}=\frac{\int d\Omega d\phi_{S_2}\int d^2{\bf q}_T(|{\bf q}_T|/M_1)\sin(\phi+\phi_{S_2})[d\sigma({\bf S}_{2T})-d\sigma(-{\bf S}_{2T})]}{\int d\Omega d \phi_{S_2}\int d^2{\bf q}_T[d\sigma({\bf S}_{2T})+d\sigma(-{\bf S}_{2T})]},
\ee
\end{widetext}
so that one easily gets
\be
\label{afinal}
\hat A_{h}   = 
-\frac{1}{2} \frac{\sum_q e_q^2
[\bar h_{1q}^{\perp(1)}(x_1) h_{1q}(x_2)+(1\leftrightarrow2)]}
{\sum_q e_q^2 [\bar f_{1q}(x_1) f_{1q}(x_2)+(1\leftrightarrow2)]}.
\ee
Thus, one can see that $\hat A_h$ is also factorized in the simple product of $\bar h_{1}^{\perp(1)}$ and $h_{1}$.

Among variety of DY processes, DY processes with antiproton ($\bar p p\rightarrow l^+l^-X$,
$\bar p p^\uparrow\rightarrow l^+l^-X$, $\bar p^\uparrow p^\uparrow\rightarrow l^+l^-X$) have essential
advantage because the charge conjugation symmetry can be applied. Indeed, due to charge
conjugation, antiquark PDF from the antiproton are equal to the respective quark PDF from the proton.
Thus, Eqs. (\ref{khat}), (\ref{afinal}) in the case of $\bar p p $ collisions are rewritten
as
\begin{widetext}
\be
\label{e16}
\hat k{\Bigl |}_{\bar pp^\uparrow\rightarrow l^+l^-X} =8\frac{\sum_qe_q^2[ h_{1q}^{\perp(1)}(x_1) h_{1q}^{\perp(1)}(x_2)+ \bar h_{1q}^{\perp(1)}(x_1) \bar h_{1q}^{\perp(1)}(x_2)]}
{\sum_qe_q^2[f_{1q}(x_1)f_{1q}(x_2)+\bar f_{1q}(x_1)\bar f_{1q}(x_2)]},
\ee
and
\be
\label{e17}
\hat A_h{\Bigl |}_{\bar pp^\uparrow\rightarrow l^+l^-X} =-\frac{1}{2}\frac{\sum_qe_q^2[ h_{1q}^{\perp(1)}(x_1) h_{1q}(x_2)+ \bar h_{1q}(x_1) \bar h_{1q}^{\perp(1)}(x_2)]}
{\sum_qe_q^2[f_{1q}(x_1)f_{1q}(x_2)+\bar f_{1q}(x_1)\bar f_{1q}(x_2)]},
\ee
\end{widetext}
where now all PDF {\it refer to protons}. Neglecting squared antiquark and  strange quark PDF
contributions to
proton and taking 
into account the quark charges and $u$ quark dominance at large\footnote{
The large $x$ values is the peculiarity of the $\bar pp$ experiments planned at GSI -- see ref. \cite{pax-tdr}}
$x$, Eqs. (\ref{e16}) and (\ref{e17}) are essentially given by 

\be
\hat k(x_1,x_2){\Bigl |}_{\bar pp^\uparrow\rightarrow l^+l^-X}  \simeq8\frac{ h_{1u}^{\perp(1)}(x_1) h_{1u}^{\perp(1)}(x_2)}
{f_{1u}(x_1)f_{1u}(x_2)},
\label{uka}
\ee
and
\be
\label{uas}
\hat A_h(x_1,x_2){\Bigl |}_{\bar pp^\uparrow\rightarrow l^+l^-X} 
\simeq -\frac{1}{2}\frac{ h_{1u}^{\perp(1)}(x_1) h_{1u}(x_2)}
{f_{1u}(x_1)f_{1u}(x_2)}.
\ee
One can see that the  system of Eqs. (\ref{uka}) and (\ref{uas}) has very simple and convenient
 form in application.
Measuring the quantity $\hat k$ in unpolarized DY (Eqs. (\ref{rhat}), (\ref{rhat1}))  and using Eq. (\ref{uka})
one can obtain the quantity $h_{1u}^{\perp(1)}$. Then, measuring SSA,  Eq. (\ref{aexpfinal}),
and using the obtained quantity $h_{1u}^{\perp(1)}$, one can eventually extract the transversity
distribution $h_{1u}$ using Eq. (\ref{uas}).
 Let us stress once again that now there is no need in any
model assumptions about ${k}_T$ dependence of  $h_1^\perp$ distributions.

In order to obtain squares of $h_{1u}^{\perp(1)}$ and $f_{1u}$ in Eqs. (\ref{uka}) and (\ref{uas}), one should consider them at the 
 points\footnote{The different points $x_F=0$ can
be reached changing $Q^2$ value at fixed $s= x_1x_2\,Q^2\equiv\tau\, Q^2$.} $x_1=x_2\equiv x$
(i.e., $x_F\equiv x_1-x_2=0$), so that
\be
\label{perp}
h^{\perp(1)}_{1u}(x)=f_{1u}(x)\sqrt{\frac{\hat k(x,x)}{8}},
\ee  
and 
\be
\label{trans}
h_{1u}(x)=-4\sqrt{2}\frac{\hat A_h(x,x)}{\sqrt{\hat k(x,x)}}f_{1u}(x).
\ee

To estimate the possibility of 
$h_{1u}^{\perp(1)}$ and $h_{1u}$ measurement, 
the special simulation of DY
events with the  PAX kinematics \cite{pax-tdr} are performed. The proton-antiproton collisions are
simulated with PYTHIA event generator \cite{pythia}. Two samples are prepared: for
the collider mode (15~GeV antiproton beam colliding on the 3.5~GeV 
proton beam) and for fixed target mode (22~GeV antiproton beam colliding on an internal hydrogen target).
 Each sample contains about 100~K
pure Drell-Yan events. Notice, that this is just the statistics planned to be achieved by PAX.
Indeed (see ref. \cite{pax-tdr}),
the sample for collider mode corresponds to about one year of data-taking
with a cross-section of
40 mb and a luminosity of $2 \times 10^{30} cm^{-2}s^{-1}$. For fixed
target mode it can takes about three months with
a cross-section of 30 mb and a luminosity of about $10^{31}cm^{-2}s^{-1}$.

Unfortunately, the original PYTHIA generator we deal with does not reproduce the corresponding to DY
experiments \cite{conway,NA10} nontrivial $q_T$ and $x$ dependencies of the quantity $\nu$ entering Eq. (\ref{cross}).
So, to estimate the 
possibility of $h_{1u}^{\perp(1)}$ and $h_{1u}$ measurement, 
one should properly introduce these dependencies in accordance with the existing experimental
data. To this end  we apply the
commonly used Monte-Carlo method  
based on weighting of the
kinematical events. 
To apply the  weighting procedure in our case, we just ascribe to each event the
weight $w=R$ which, in accordance with the data \cite{conway,NA10},
is given by Eq. (\ref{cross}),
where $\lambda\simeq1$, $\mu\simeq0$ and $\nu$ has nontrivial $q_T$ and $x$ dependencies.
The $q_T$ dependence of $\nu$ is taken from refs. \cite{bo1,bianconi}  -- Eq. (49) in ref. \cite{bo1} and Eq. (21) in ref. \cite{bianconi},
and this $q_T$ dependence properly fits the existing experimental data \cite{conway,NA10}.
However, in refs. \cite{bo1,bianconi}
 (where the simplified Boer's model is applied) there is no (important and corresponding 
to DY experiments \cite{conway,NA10})
 $x$-dependence of $\nu$ at all, so that we take this dependence from ref. \cite{conway}.

To check the validity of the angular distribution analysis of the weighted events 
we reconstruct the ${q_T}$ and $x_1$ dependencies of $\nu$. 
The results are shown  in Figs. 1, 2.
\begin{figure}[h!]
\includegraphics[height=5cm,width=8cm]{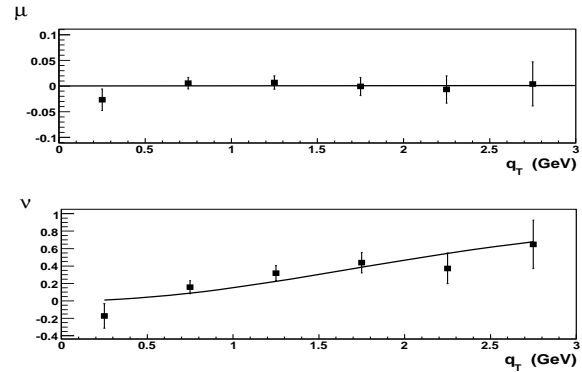}
\caption{\it 
Reconstructed from simulations (fixed target mode)  quantities 
$\mu$ and $\nu$ versus $q_T$ in comparison
with the  input (corresponding to experimental data) dependencies (solid lines).
 }
\label{pic2}       
\end{figure}
\begin{figure}[h!]
 \includegraphics[height=5cm,width=8cm]{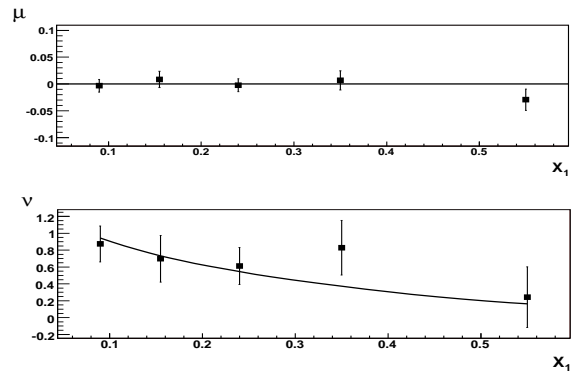}
\caption{\it  
Reconstructed from simulations (fixed target mode)  quantities 
$\mu$ and $\nu$ versus $x_1$ in comparison
with the  input (corresponding to experimental data) dependencies (solid lines).}
\label{pic3}       
\end{figure}
One can see a good agreement\footnote{As an additional check of our analysis validity,
 we  reproduce the input zero value of $\mu$.} between input (solid lines) and
reconstructed (points with error bars) values. 

\begin{figure}[ht!]
\begin{center}
\includegraphics[height=5cm,width=8cm]{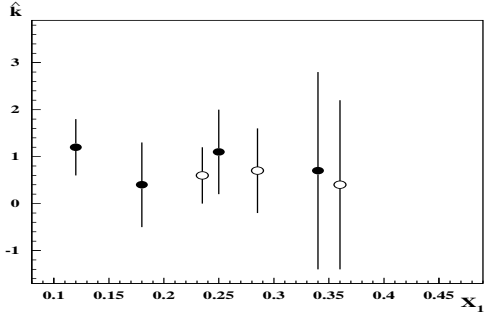}
\end{center}
\caption{\it $\hat k$  versus $x_1$ at $x_F\simeq0$. Data is obtained with MC simulations for
collider (closed circles) and for fixed target mode (open circles). 
For better visibility (to avoid overlapping) the points for collider (fixed target)
mode are shifted 0.01 to the left (right) along the x-axis. }
\label{k-hat}       
\end{figure}

Thus, applying the above described weighting procedure,
our simulations reproduce the nontrivial  angular dependence of $R$  with  $q_T$- and
$x$-dependent $\nu$.
These dependencies are in accordance with the respective
input dependencies obtained in experiments on DY \cite{conway,NA10}.
Now it is straightforward to reconstruct  the $q_T$-weighted quantity  $\hat R$
(Eq. \ref{rhat}) and, consequently, $\hat k$ (Eq. \ref{rhat1}).
The results are shown in Fig. \ref{k-hat}. 
The values of $\hat k$ at averaged Q$^2$ for both modes are found to be $1.2 \pm 0.2$ for
collider mode and  $1.0 \pm 0.2$ for fixed target mode. 

\begin{figure}[t!]
\begin{center}
\includegraphics[height=5cm,width=8cm]{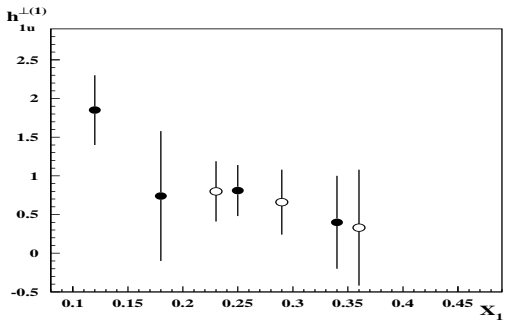}
\end{center}
\caption{\it $h_{1u}^{\perp(1)}$  versus $x_1$ at $x_F\simeq0$. Data is obtained with MC simulations for
collider (closed circles) and for fixed target mode (open circles). For better visibility (to avoid overlapping) 
the points for collider (fixed target)
                mode are shifted 0.01 to the left (right) along the x-axis.}
\label{hp-hat}       
\end{figure}

The quantity $h_{1u}^{\perp (1)}$ is reconstructed 
from the obtained values of $\hat k$
using Eq. (\ref{perp}) with $x_F=0 \pm 0.04$. The results are shown in Fig. \ref{hp-hat}.
The obtained magnitudes 
of $h_{1u}^{\perp (1)}$ are in accordance (in order of value) with the respective magnitudes 
obtained with the model (\ref{bmodel}) for  $h_{1u}^{\perp}(x,{\bf k}_T)$. Indeed, for example for
the collider mode ($Q_{average}^2 \simeq 9~GeV^2$, so that $x_1\simeq x_2 \simeq 0.2 $
at the point $x_F\simeq 0$) the results from the simulations and from the  model (\ref{bmodel}) are
$h_{1u}^{\perp (1)} \simeq 1$ and  $h_{1u}^{\perp (1)} \simeq 0.5$, respectively.

Using the obtained magnitudes of $h_{1u}^{\perp (1)}$ we estimate the expected SSA given by Eq. (\ref{uas}).
The results are shown in Figs. 5 and 6. 
For estimation of $h_{1u}$ entering SSA together with $h_{1u}^{\perp (1)}$ (see Eq. (\ref{uas})) we follow the procedure of ref. \cite{anselmino1} 
and use (rather crude) ``evolution model''
\cite{bo1, anselmino1} , where there is no any estimations of uncertainties. That is why in (purely 
qualitative) figures 5 and 6 we present the solid curves instead of points with error bars.
To obtain these curves we reproduce  $x$-dependence of $h_{1u}^{\perp(1)}$  in the considered region, 
using the Boer's model, Eq. (\ref{bmodel}), 
properly numerically corrected in accordance with the simulation results.

To estimate the measurability of the quantities we deal with,
it  is relevant to estimate the upper bounds 
on   $h_1$, $h_1^{\perp(1)}$ and then on $\hat k$ and $\hat A_h$. 
Obtaining $h_{1u}^{\perp(1)}$ and $h_{1u}$ one deals
with Eqs. (\ref{uka}), (\ref{uas}) applied at the points $x_1\simeq x_2\simeq\sqrt{Q^2/s}$,
so that we perform the estimation of the upper bounds on $\hat k$ and $\hat A_h$ at the points $x_F\simeq0$
corresponding to the average $Q^2$ values for both collider and fixed target modes.
The maximally allowed value of $h_{1}^{\perp(1)}$ can be found operating just as it was
done with respect to the quantity $f_{1T}^{\perp(1)q}$ (first moment of the Sivers function)
in ref. \cite{efremov}. To this end we first apply the inequality\footnote{This inequality 
is directly obtained by relaxing the bound Eq. (16) in ref. \cite{mulders-ineq} (eliminating the unknown distribution in that bound).}
\cite{mulders-ineq} 
$
({|{\bf k}_T|}/{M})h_{1}^{\perp}(x,{\bf k}_T^2)\le f_1(x,{\bf k}_T^2).
$
Then, using the estimation (see ref. \cite{efremov} and references therein) $\langle k_T\rangle\simeq0.8\,GeV$
one easily gets the upper bound on $h_{1u}^{\perp(1)}$:
$
h_{1u}^{\perp(1)}\simleq0.4f_{1u}(x).
$
On the other hand, maximally allowed value of $h_{1u}$ can be found using the Soffer \cite{soffer} inequality
$
|h_{1u}|\le (f_{1u}+g_{1u})/2
$. For the PAX kinematics $s=43\,GeV^2$, $Q^2_{average}\simeq5\,GeV^2$  for the fixed target mode and
$s=215\, GeV^2$, $Q^2_{average}\simeq 9\, GeV^2$ for the collider mode. Thus, at the point $x_F=0$ we deal with, $x_1\simeq x_2\simeq 0.3$
 and $x_1\simeq x_2\simeq 0.2$
for the
fixed target and collider modes, respectively.
Then, the inequalities on $h_{1u}$ and $h_{1u}^{\perp(1)}$ give\footnote{
Performing these estimations we use GRSV2000LO parametrization \cite{grsv} for $g_{1u}$ and 
GRV98LO parametrization \cite{grv98} for $f_{1u}$.
} $h_{1u \,(max)}\simeq 1.5$ ($f_{1u}=1.9$, $g_{1u}=1.0$ )
and $ h_{1u\,(max)}^{\perp(1)}\simeq 0.8$ for fixed target mode 
while $h_{1u\,(max)} \simeq 2.3$ ($f_{1u}=3.1$, $g_{1u}=1.5$) and $h_{1u\,(max)}^{\perp(1)}\simeq1.2$
for collider mode. 
Using these estimations of $h_{1u\,(max)}$ and  $h_{1u\,(max)}^{\perp(1)}$ 
in Eqs. (\ref{uka}), (\ref{uas})  
it is straightforward to obtain the maximally allowed values of $\hat k$ and $\hat A_h$: 
${\hat k}_{(max)}\simeq1.4$ and $|{\hat A}_{h\,(max)}|\simeq0.17$ for fixed target mode while
${\hat k}_{(max)}\simeq1.2$ and $|{\hat A}_{h\,(max)}|\simeq0.14$ for collider mode.
One can see that obtained estimations of upper bounds on $h_{1u}^{\perp(1)}$, 
$\hat k$ and $\hat A_h$ are in accordance
with the results presented by Figs. 3-6.

Looking at the (preliminary) estimations presented by Figs. 3 and 4, one can conclude
that the quantities $\hat k$ and $h_{1u}^{\perp(1)}$ are presumably measurable in  most 
of the considered $x$-region. 
At the same time, looking at Figs. 5 and 6 one can see that for 
both modes SSA $\hat A_h$  is estimated to be about 6-8\%.
On the other hand, as it was argued in Ref. \cite{pax-tdr} (see section ``Single Spin asymmetries and Sivers Function'', p. 25),
the  studied in ref. \cite{efremov} SSA $A_{UT}^{\sin(\phi-\phi_S)\frac{q_T}{M_N}}$ of order 5-10\% can be measured by PAX.
It is obvious that studied in this paper SSA $\hat A_h$, weighted with $\sin(\phi+\phi_S)$ and the same weight $q_T/M_N$, 
is absolutely analogous to SSA  $A_{UT}^{\sin(\phi-\phi_S)\frac{q_T}{M_N}}$,
so that it is clear that if  $A_{UT}^{\sin(\phi-\phi_S)\frac{q_T}{M_N}}$ of 5-10\% is measurable, then $\hat A_h$ of 6-8\% is measurable too.

\begin{figure}[t!]
\begin{center}
\includegraphics[height=5cm,width=8cm]{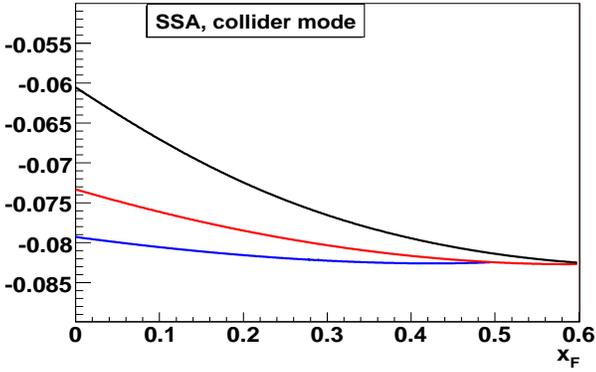}
\end{center}
\caption{ \it SSA given by Eq. (\ref{uas}) versus $x_F$ for collider mode for three values of Q$^2$: 50~GeV$^2$ (lower curve),
25~GeV$^2$ (middle curve) and 9~GeV$^2$ (upper curve).}
\label{ssa-collider}       
\end{figure}

\begin{figure}[t!]
\begin{center}
\includegraphics[height=5cm,width=8cm]{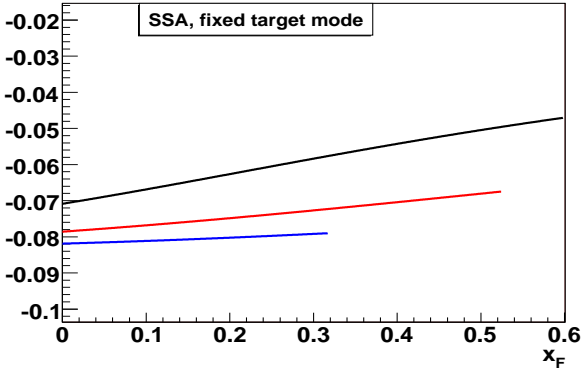}
\end{center}
\caption{ \it SSA given by Eq. (\ref{uas}) versus $x_F$ for fixed target mode for three values of Q$^2$: 16~GeV$^2$ (lower curve),
9~GeV$^2$ (middle curve) and 4~GeV$^2$ (upper curve).}
\label{ssa-fixed}       
\end{figure}

Thus, it is shown that it  is possible to directly extract the transversity and its accompanying
T-odd PDF from the unpolarized and single polarized DY processes with antiproton participation.
It is of importance that there is no need in any model assumptions about ${k}_{T}$ dependence of
$h_1^{\perp}$.
 One can directly extract both $h_1$ and first moment of $h_1^{\perp}$ from the single-polarized and
unpolarized DY processes, since these quantities enter the measured $\hat k$ and SSA $A_h$ in the form of simple
product instead of complex convolution.     
The preliminary estimations for PAX kinematics show the possibility  to measure both $\hat k$ and SSA $\hat A_h$ and
then to extract  the quantities $h_1^{\perp (1)}$ and  $h_1$.
Certainly, the estimations 
of $\hat k$ and $\hat A_h$ magnitudes obtained it this paper
are very preliminary and show just the order of values of these quantities.
For more precise estimations one needs the Monte-Carlo generator   
more suitable for DY processes studies (see, for example. ref. \cite{bianconi} ) than PYTHIA generator which we used 
(with the proper weighting of events) here.

Notice, that it is straightforward to properly modify the procedure discussed in this paper
to DY processes: $\pi^-p \to \ \mu^+ \mu^- X$ and
$\pi^-p^{\uparrow} \to \ \mu^+ \mu^-  X$, which could be study \cite{torino}
in the COMPASS experiment at CERN.

  The authors are grateful to M.~Anselmino, R.~Bertini,
 O.~Denisov, A.~Efremov, A. Kacharava, V.~Krivokhizhin,
 A.~Kulikov, P.~Lenisa, A.~Maggiora, A.~Olshevsky, 
G.~Piragino, G.~Pontecorvo, F.~Rathmann, I.~Savin, M.~Tabidze, O.~Teryaev and W. Vogelsang
 for fruitful discussions. The work  of O.S. and O.I. was supported by the Russian Foundation
 for Basic Research (project no. 05-02-17748).

\end{document}